# Anisotropic Metal-Insulator Transition in Epitaxial Thin Films

I. B. Altfeder[1,2*], X. Liang[2,3], D. M. Chen[2], and V. Narayanamurti[1]

[1]Division of Engineering and Applied Sciences, Harvard University, Cambridge, MA 02138

[2]The Rowland Institute at Harvard, Cambridge, MA 02142

[3]Institute of Physics, Chinese Academy of Sciences, Beijing, 100080

Quantum wells made of simple polyvalent metals represent a novel family of doped 2D Mott-Hubbard insulators. As scanning tunneling microscopy experiments show, these systems exhibit an anisotropic form of metal-insulator transition. Their elementary excitations possess coherent wave-like properties along the normal axis, and show an incoherent behavior in-plane. The development of such an anisotropic coherence is most likely related to Coulomb interaction between localized and delocalized thin film electronic states – 2D Kondo screening.

Quantum enhancement of strongly correlated phenomena represents an important motivation for the experimental study of nanostructures[1-3]. The evidence of unusual electronic localization in atomically flat thin Pb(111) films has previously been shown in scanning tunneling microscopy (STM) experiments[4-6]. A simple quasiclassical model has been proposed for the explanation of this effect. Owing to nested hole Fermi surface, the electrons tunneling into these films move only in the direction of $k$-quantization, that causes the formation of quantum dot-like (0D) states. The alternative manifestation of this effect, where the Fermi surface nesting and the transport anisotropy are *induced* by thin film geometry, represents yet an unsolved challenge.

In principle, non-equilibrium 0D states may arise not only from the material anisotropy, but also from the enhanced many-body interactions in a thin film, facilitated by finite-size/reduced dimensionality[2, 7, 8]. Indeed, considering these states as "electronic impurities"[9] one would expect the manifestation of significant spin/charge effects, including Kondo screening. Experimentally, the existence of such correlations can be revealed by replacing Pb with another material, whose bulk Fermi surface is more free-electron-like. The good candidate for this experiment would be a metal having one *p*-electron per atom less than Pb, as for example In[10, 11]. For the STM experiment, a thin In film represents a perfect comparison object for yet another reason: room-temperature deposition of both Pb and In on Si(111) yields nanostructures of very similar geometry, i.e. (111)-oriented flat-top islands with the face-centered cubic lattice[12, 13]. Moreover, in both cases the 7x7 reconstruction of Si(111) remains unchanged upon burial. Due to geometric similarity of these nanostructures, the difference of their electronic properties will originate primarily from the difference of valence numbers (quasi-doping effect) and Fermi surface topologies.

Here, we present an in-depth comparison of the electronic properties of thin Pb(111) and In(111) films. Our study suggests that thin epitaxial films of simple polyvalent metals represent a family of doped 2D Mott-Hubbard insulators, whose anisotropic transport behavior is most likely stabilized due to Coulomb interaction between localized and delocalized electronic states.

The experiments were performed in a dual-chamber ultra high vacuum (UHV) system equipped with surface preparation and analysis tools and with a low-temperature STM. A silicon substrate was cleaned by a sequence of flashing to 1100°C and ion sputtering. After a clean 7x7 reconstructed Si(111) surface was obtained, metal atoms were deposited from an effusion cell with a rate of ~1 monolayer (ML)/min to form an epitaxial film. The quality and the chemical composition of the surface both before and after deposition were monitored using reflective high-energy electron diffraction and Auger spectroscopy. The sample was then transferred *in-situ* into the STM analysis chamber.

In Fig. 1, we show a typical STM image of In(111) film obtained at 77 K. Following the Stranski-Krastanov mode, on top of the In wetting layers grow 3D flat-top nanocrystals. Careful analysis shows that the majority of islands are confined between the substrate steps, forming sequences oriented along the step edges. In Fig. 1, two of such sequences are indicated by dashed lines. This effect is probably caused by a diffusion barrier at a step edge, which confines deposited atoms within a single terrace. Anisotropic diffusion resulting from such a confinement may be responsible for elongated island geometry. The analysis of height distribution of the islands reveals forbidden and allowed values, resembling the "electronic growth" due to $\lambda_F / 2$ transverse modulation of charge density, reported earlier for thin Pb films[14]. Since for In(111) $\lambda_F \approx 8a$,[10] the expected periodicity of allowed heights ~ 4 ML.

In Fig. 2a, we show the tunnel density-of-states spectra, measured for different In islands (heights ≤ 15 ML) found in our samples. The spectra exhibit distinct resonant features, whose energy separation decreases with the increase of thickness. Such a behavior is expected for electron confinement in a one-dimensional quantum box, where quantized states, $E_n$, are separated by a gap $\Delta = \pi \hbar v / H$, with v being the electron velocity, and $H$- film thickness. The change of tunnel spectra with the decrease of metal height by 1 ML is illustrated in Fig. 2c. As was shown earlier[15], 1 ML thickness change introduces a spectral shift $\delta \approx \Delta (2a / \lambda_F)$. In the case of In(111) films, we found $\delta \approx \Delta / 4$. The narrow, $\Sigma \approx 0.4 \, eV$, width of the 2D subbands indicates a large 2D effective mass, not expected for a free-electron metal. The spectral data are summarized in Fig. 2b, where we plot $\Delta^{-1}$ as a function of the number of atomic layers. The electron velocity perpendicular to the interface, as determined from Fig. 2b, $v_\perp \approx 2 \times 10^8 \, cm/sec$. The thickness-offset of the film, introduced by In wetting layers, may be responsible for the horizontal offset in Fig. 2b.

Thus, the twofold rescaling of quantum effects with the change of electron wavelength from $\lambda_F^{Pb} \approx 4a$ to $\lambda_F^{In} \approx 8a$ is clearly manifested along the axis [111]. As we shall show later, a similar wave-like process does not take place in the (111) plane.

Owing to [111]-directed coherent electron transport, the STM images of In(111) islands reveal information about the buried In/Si(111) interface. In Figs. 3 (a-d), we show the sequence of voltage-dependent STM images obtained on top of 7 ML ($H = 9a$) island. Comparison of the images indicates partial preservation of the 7x7 reconstruction, accompanied by self-assembly of a nanoscale domain structure due to interfacial alloying. The bright defects, seen in Fig. 3a ($V_{tip} = 250 \, mV$), illustrate the "tops" of subsurface pyramid-shaped structures, created by alloying of Si and In atoms. The change of tunnel bias in Fig. 3b ($V_{tip} = -250 \, mV$) shifts the vertical position of the imaging plane $h = \pi n / k_\perp(E)$ closer to the interface. Now, the 7x7 reconstructed In/Si(111) interface becomes visible in the STM image, while the bright defects increase in size and transform into nanoscale islands. Further increase of the tunnel bias (Fig. 3c, $V_{tip} = -450 \, mV$) additionally enhances the contrast of the interfacial image, making the area occupied by islands even larger. In a situation of Fig. 3c, the imaging plane coincides with the interface ($h = H$), and STM resolves the "bases" of three-dimensional interfacial structures, rather than their "tops" (seen in Fig. 3a). The electronic image of the buried interface disappears in Fig. 3d due to off-resonance situation ($V_{tip} = -850 \, mV$).

Quite surprisingly, despite free-electron character of bulk In Fermi surface (see Fig. 3b') the structure of buried In/Si(111) interface can be resolved in fine details. The STM cross-sections, shown in Fig. 3a', suggest the lateral resolution $\xi \approx 0.7 \, nm$. Moreover, the STM images do not reveal any signature of lateral interference fringes around subsurface defects[16], whose formation is expected for coherent 3D Bloch waves. On the contrary, such an anisotropic one-dimensional character of interference would be typical for the array of incoherently coupled ([111]-*oriented*) one-dimensional channels[17, 18]. As the bulk material does not exhibit a flat Fermi surface (large 2D effective mass), its appearance is likely to be caused by thin-film geometry.

The difference of electronic configurations of In and Pb islands manifests not only in the twofold change of $\lambda_F$. As we found from the analysis of broad energy range spectroscopic measurements, the change of electron density can also be seen in tunnel spectra. In the upper inset of Fig. 4, we show the current-voltage characteristic of In island. The resonant feature at −0.6 $V$ is clearly resolved. Applying a large (>1 $eV$) tunnel bias significantly modifies the shape of a potential barrier, resulting in an exponential enhancement of the tunnel matrix elements and a strong *I-V* nonlinearity. This non-linearity can be eliminated in the normalized *I-V* spectra, shown in Fig. 4, where the data were measured in a broad (± 4 $V$) range of tunnel bias and presented using a linear-logarithmic vertical scale *arcsinh(I)*. In addition, in Fig. 4 we show the normalized *I-V* spectrum of a Pb(111) sample, probed in the same range of tunnel bias. Comparison of the curves shows, that while in In(111) films the quantum dot-like interference persists to the very top of a potential well (with little $\sum_n$-dependence), in Pb(111) films such an interference abruptly vanishes when the energy of tunneling electrons exceeds 2.0-2.5 $eV$. The resonant behavior of STM images, similar to shown in Figs. 3(a-d), has been observed across the corresponding range of tunnel bias. It seems natural to suggest that the rise in energy of resonances in In(111) films originates from the reduction of the Fermi energy as a consequence of reduced electron density.

In the tight-binding model, electronic states near the middle of the conduction band (half-filling) form a quasi-1D gas with linear dispersion $E(\mathbf{k}) = v_f \hbar (k_\perp - \pi/2a)$. The next order quadratic terms are absent since in the middle of the band the effective masses become infinitely large. Confinement of these quasi-1D electrons in a finite-size potential well creates 0D states with equidistant energy levels ($E_n$). In principle, moving away from the middle of the conduction band introduces a non-zero lateral bandwidth ($\varepsilon_n$). Following the argumentation of metal-insulator transition, the transformation of 0D-states into 2D-states occurs when $\varepsilon_n > U$, where $U$ is the on-site "bound" energy. According to Fig. 4, quantum dot-like states develop inside $2U \approx$ 4-5 $eV$ energy window, whose center is pinned in the middle of the conduction band. For a metal with almost half-filled band, such as Pb, this energy window is symmetric around the Fermi level, as it is illustrated in the lower inset (a) in Fig. 4. For In, whose conduction band is significantly less than half-filled, the energy window for 0D states becomes asymmetrically lifted with respect to the Fermi level, as shown in the lower inset (b) of Fig. 4. Since tunnel spectroscopy primarily detects unoccupied 0D states[19], only unoccupied portions of these energy windows can be seen in our data in Fig. 4. Thus, a virtual "depletion" of the conduction band ($s^2 p^2 \rightarrow s^2 p^1$) lifted above the Fermi level a part of the occupied resonant states.

A possible reason for the 2D anomalies observed in our samples, including (a) the appearance of large 2D effective mass, (b) the absence of lateral interference, and (c) the signature of 2D bound energy, could be the formation of 2D Kondo/Anderson lattice due to Coulomb interaction between *localized* and *delocalized* states[20] (see Fig. 3c'). In a simplest description of this effect, the electrons sequentially injected into a quantum well become surrounded by a cloud of antiparallel spins. This spin geometry is known to produce an additional Coulomb repulsion (*correlation*) energy $U$ that "confines" electrons at a given point of (111)-plane, apparently not affecting their ability to perform transverse ballistic oscillations (2D Kondo screening). Through such a "confinement", coherent 2D Bloch waves transform into incoherent heavy electrons, shown in Fig 3c'.

In conclusion, a surprising interplay between the transverse and the lateral electron confinement in a thin film, as discussed in this Letter, represents a clear example of *quantum enhancement* of Coulomb interaction.[21] This enhancement, unlike in previously known models, arises from the lack of a microscopic disorder, i.e. from the elastic electron reflections at atomically flat crystal boundaries. Indeed, absent for transverse confinement ($H = \infty$) the screening cloud is unable to follow the electron, and the lateral bound states do not form. The family of doped 2D Mott-Hubbard insulators, arising from the anisotropic thin-film interactions, may also include more complicated layered structures exhibiting "intermediate" electronic configurations $s^2 p^x$.

The authors acknowledge F. Spaepen, L. P. Pryadko, and A. V. Chubukov for interesting discussions. The research was supported by NASA Ames Laboratory and Harvard NSF funded Nanoscale Science and Engineering center (NSEC).

\* altfeder@deas.harvard.edu

1. O. Pietzsch, A. Kubetzka, M. Bode, *et al*., Science **292**, 2053 (2001).

2. D. J. Huang, G. Reisfeld, and M. Strongin, Phys. Rev. B **55**, R1977 (1997).

3. B. G. Orr, H. M. Jaeger, and A. M. Goldman, Phys. Rev. Lett. **53**, 2046 (1984).

4. I. B. Altfeder, V. Narayanamurti, and D. M. Chen, Phys. Rev. Lett. **88**, 206801 (2002).

5. W. B. Su, S. H. Chang, W. B. Jian, *et al*., Phys. Rev. Lett. **86**, 5116 (2001).

6. I. B. Altfeder, D. M. Chen, and K. A. Matveev, Phys. Rev. Lett. **80**, 4895 (1998).

7. R. Egger and H. Grabert, Phys. Rev. Lett. **79**, 3463 (1997).

8. J. van den Brink and G. A. Sawatzky, Europhys. Lett. **50**, 447 (2000).

9. D. Goldhaber-Gordon, H. Shtrikman, D. Mahalu, *et al*., Nature **391**, 156 (1998).

10. N. W. Ashcroft and W. E. Lawrence, Phys. Rev. **175**, 938 (1968).

11. J. R. Anderson and A. V. Gold, Phys. Rev. **139**, A 1459 (1965).

12. A. Pavlovska, E. Bauer, and M. Giessen, J. Vac. Sci. & Techn. B **20**, 2478 (2002).

13. M. Yoon, X. F. Lin, I. Chizhov, *et al*., Phys. Rev. B **64**, 085321 (2001).

14. M. Hupalo and M. C. Tringides, Phys. Rev. B **65**, 115406 (2002).

15. I. B. Altfeder, K. A. Matveev, and D. M. Chen, Phys. Rev. Lett. **78**, 2815 (1997).

16. M. F. Crommie, C. P. Lutz, and D. M. Eigler, Nature **363**, 524 (1993).

17. P. W. Anderson, Phys. Rev. Lett. **67**, 3844 (1991).

18. V. J. Emery, E. Fradkin, S. A. Kivelson, *et al*., Phys. Rev. Lett. **85**, 2160 (2000).

19. At a positive tip voltage, the transport is pinned at the highest occupied quantum state, making observation of deeper occupied states impossible.

20. Apparently, only (111)-segment of hole Fermi surface can be affected by localization.

21. Although not for thin films, a transformation of a free-electron metal into an insulator has been discussed in theory: A. W. Overhauser, Phys. Rev. Lett. 4, 462 (1960); A. W. Overhauser and L. L. Daemen, Phys. Rev. Lett. 61, 1885 (1988). We therefore cannot exclude a possible connection between these works and our results.

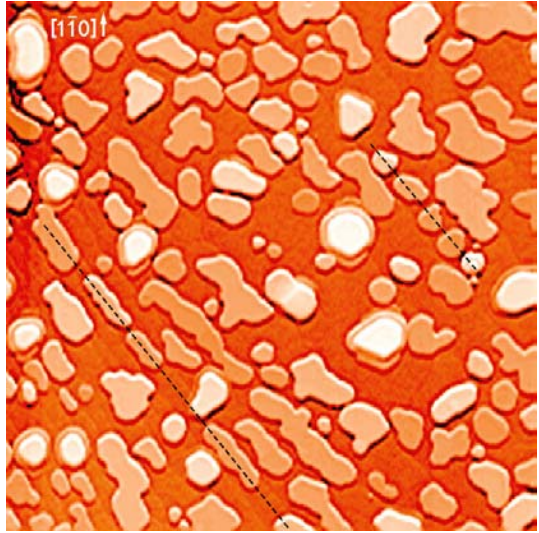

**Figure 1 (color)**

A typical STM image of In islands on Si(111). The image size is $1600 \times 1600$ nm$^2$.

Altfeder et al. Fig.1 (color)

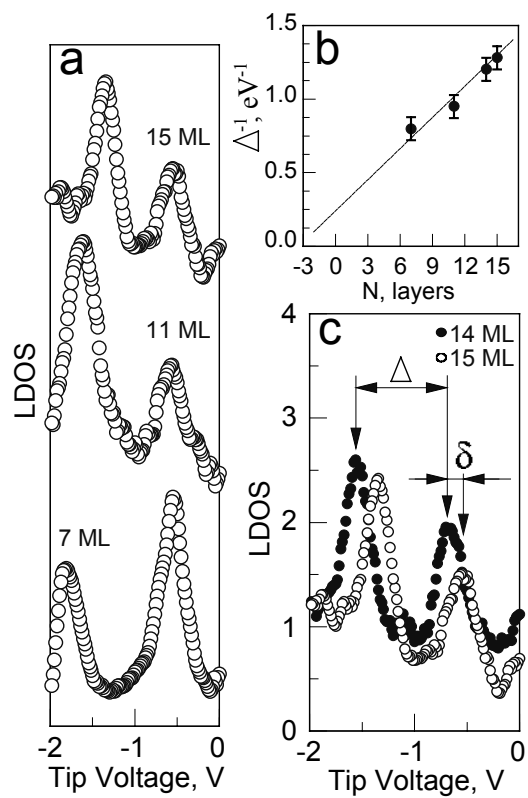

## Figure 2

(a) The density-of-states spectra for different In islands.

(b) The dependence of $\Delta^{-1}$ on the number of atomic layers.

(c) Spectral shift introduced by 1 ML thickness change in a wedge-shaped island.

Altfeder et al. Fig. 2

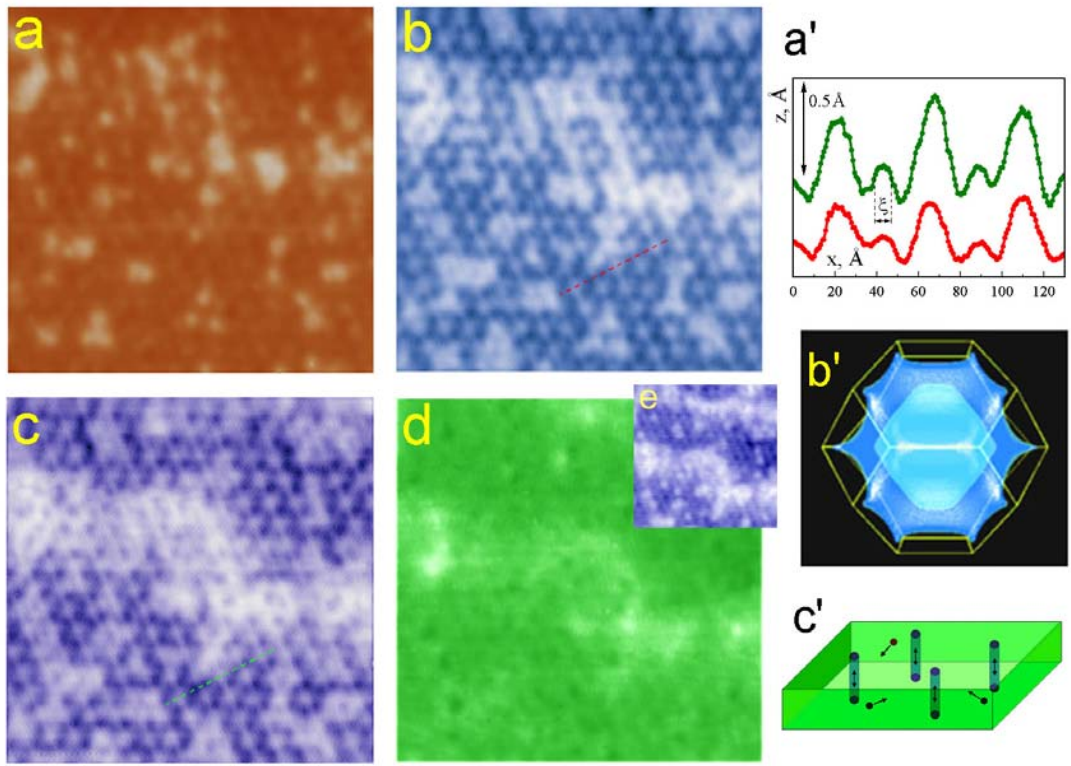

**Figure 3 (color)**

(a-d) The $400 \times 400$ Å$^2$ STM images of 7 ML island, obtained at different tip voltages:

(a) $V=250$ $mV$,  (b) $V=-250$ $mV$, (c) $V=-450$ $mV$, (d) $V=-850$ $mV$.

(e) Reappearance of a resonant image at $-3.4$ $V$.

(a′) The cross-sections of Figs. 3(b-c), where large and small peaks originate from subsurface corner-holes and dimer dislocations.

(b') Hole Fermi surfaces of In and Pb (inner).

(c') Anisotropic Mott-Hubbard excitations (vertical columns) and Fermi liquid quasiparticles.

Altfeder et al. Fig. 3 (color)

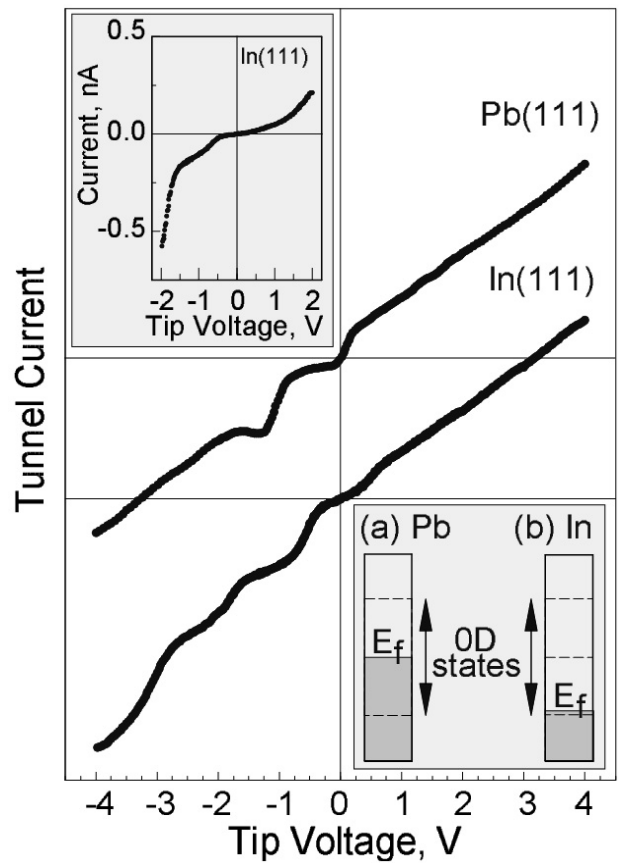

**Figure 4**

Normalized tunnel *I-V* characteristics of In and Pb films, obtained in ±4 *V* bias range.

Upper inset: tunnel *I-V* characteristic of In island in a bias range ±2 *V*.

Lower inset: the energy window of 0D-states is pinned in the middle of the conduction band: (a)

half-filled band, (b) less than half-filled band.

Altfeder et al. Fig. 4